Piotr Guzowski
University of Bialystok
Faculty of History and International Relations
Center for the Study of Demographic and Economic Structures in Preindustrial Central and Eastern Europe
Plac NZS 1
15-420 Bialystok
Poland
guzowski@uwb.edu.pl

**Did the Black Death reach the Kingdom of Poland in the middle of the 14<sup>th</sup> century?**

**Abstract**

The Black Death is regarded as a turning point in late medieval European history. Recent studies have shown that even regions that have so far been perceived in the literature as not or only marginally affected by the epidemic, suffered from its profound demographic and economic consequences. The scale and geographical range of the plague in Central Europe, the Kingdom of Poland included remains, however, a matter of dispute and from the beginning scholars' views on this matter have been divided. What is particularly important, the outbreak of the plague in Western Europe coincided with the reign of Casimir of the Piast dynasty, the only ruler of Poland to receive the nickname the Great, who is associated with the modernisation and extraordinary development of his kingdom which was entering the golden age of its history.

**Keywords: Black Death, Kingdom of Poland, Cracow, Peter's Pence**

The Black Death is considered a turning point in the social and economic history of Europe and for almost two hundred years it has been of interest to historians of various specializations[i]. In recent years, thanks to the interdisciplinary approach to the mid-14th century pandemic, a number of issues related to the biological aspects of the causes and spreading of the disease have been addressed. Its nature and origins have been studied[2], and its importance in the context of environmental history has been emphasized. Over the last few decades, geography and chronology of the Black Death have also been clarified. In Richard C. Hoffmman's latest study, the areas of the Kingdom of Poland and the Czech Kingdom have been defined as the only, apart from some territories in the Pyrenees, areas of low plague mortality.[3]. The aim of this article is to answer the question whether the first wave of the Black Death reached Poland in the mid 14th century and whether it influenced the economic and demographic situation of the country undergoing modernization under the rule of Casimir, the only king in Polish history bearing the nickname the Great. The relevance of this question transcends the Central European contexts of studies in the Black Death. It provides an opportunity to test the validity of epidemic spreading models that assume similar development of the disease and comparable population losses in all European regions, even those where the spread of the epidemic cannot be quantified due to the scarcity of sources.[4] The article aims to shed some light on the specifically Central European interpretation of the consequences of the Black Death, but it also contextualises the analysis of Polish sources and indicates how it can contribute to the debates on the scale and nature of late medieval economic crisis, the paths of European economic development and the emergence of what is known as Little Divergence.[5]

**Historiographical view**

Among scholars studying the Black Death, the positions on whether the Kingdom of Poland was affected by the plague have always been divided. The author of the earliest important monograph on this subject, Robert Hoeniger, claimed that medieval Poland was free of the plague. Meanwhile, Georg Sticker takes an opposing view and provides a description of the outbreak of the epidemic in Pomeranian cities belonging to the Teutonic Order and in Silesia, that is in lands bordering on the Kingdom of Poland, and in the Kingdom itself.[6]

Polish historiography has not produced a definite stand on the subject of the Black Death. On one hand, the findings of Piotr Rutkowski, Jan Tyszkiewicz, and Jerzy Janowski in the field of the history of medicine suggest that the plague was indeed present in the Kingdom of Poland. The authors of popular science studies are of a similar opinion.[7] On the other, the authors of works on medieval Poland's history and economy emphasize that while the plague was destroying the structures of

Western European societies and economies, the Polish Kingdom thrived under the rule of king Casimir. He earned his title, the Great, which he earned as a modernizer of Polish economy and institutions of the state[8]. It is also emphasized that although the plague was present in Teutonic Order's lands north of the Polish border, it must not imply its presence in the Kingdom itself.[9] According to contemporary research, there is no evidence that the disease reached Silesia either, in spite of the appearance of the flagellants, who were popularly associated with the panics caused by the outbreaks of the epidemic in Central Europe.[10] Difficulties in presenting any conclusive arguments to support or disprove the hypothesis of the arrival of the first wave of the Black Death in Polish lands in the later Middle Ages have forced historians to formulate evasive, vague statements whenever they wrote about the population of this area, such as: "Demographic development was observed, despite periodical, devastating epidemics, which affected also towns and cities".[11] Also in the ongoing and inconclusive discussion on the issue of the economic crisis of the late Middle Ages in Poland, the argument about possible depopulation caused by the Black Death epidemic is not raised.[12]

Authors of the publications written since the middle of the 20th century also represent two contrasting views on the question of whether or not the epidemic reached Poland. The "absence" view is best represented by Élizabeth Carpentier, who in her article in *Annales*, entitled *Autour de la peste noire: famines et épidémies dans l'histoire du XIVe siècle* provides a map showing the Kingdom of Poland as unaffected by the Black Death.[13] Robert S. Gotfried, on the other hand, suggests that in Central Europe, the epidemic resulted in a lower population loss than in the West and estimates the decline of the Polish population at 25%.[14] Meanwhile, Ole J. Benedictow presents a different view in his two studies on the Black Death in Europe[15]. He claims that in the light of the epidemiological theory describing the mechanism behind the contagion, it is highly unlikely that the Kingdom of Poland and the Czech Kingdom were free of the Black Death.[16] He assumes that since the pestilence was present in Germany, Prussia and Russia, Poland could not have escaped it either. Benedictow suspects that historians in communist Poland did not undertake any research on this subject for political reasons. A Marxist approach that they were expected to apply to all their studies did not allow them to propose the epidemic as a factor alternative to class struggle to explain profound economic changes and the creation of a socio-economic system based on demesne and second serfdom. In his first book, relying heavily on the findings of older German historiography, and in the later one, in which Polish primary sources are also included, Benedictow makes a claim that the plague spread throughout both Polish Kingdom and Kingdom of Bohemia.

Benedictow's hypothesis presented in 2004 encouraged scholars studying medieval Bohemia and Poland to engage in discussions about the Black Death. The views that prevailed in earlier Czech historiography were revised by D.C. Mengel in an article published in *Past and Present*, in which the author developed a general critique of Benediktow's claims, but at the same time he did not exclude the possibility of the presence of the plague in Bohemia in the mid 14th century.[17] Benedictow responded to the critique in the book published in 2021, in which he maintained his previous claim that the Black Death did reach mid-14th-century Bohemia[18]. In the case of Poland, indirect data on mortality among the highest Church and state dignitaries in the middle of the 14th century as well as data concerning the foundations of churches and altars were compiled by a group of Polish historians and archaeologists. Unfortunately, the study did not yield any clear-cut answers to the question of whether or not the Black Death reached Poland.[19] In the latest Western European historiography, authors writing about other parts of the continent have made occasional references to Poland, claiming that it might have been affected by the plague, or have mentioned it among the countries whose population decreased in the middle of the 14th century.[20] However, these assertions have not been supported by references to Polish primary or secondary sources. Nevertheless, analyses of narrative sources have recently led Polish historians to believe that the theory of the outbreak of the first wave of the Black Death in Poland is plausible.[21] In his 2021 work, Benedictow bases his argumentation on the analysis of late medieval and early modern narrative sources as well as the observation of the fluctuations of prices in late medieval Cracow. Changes in prices, as he claims, provide evidence to support the hypothesis of the demographic crisis in the Polish lands. In addition to responding to Benedictow, in its individual sections this article describes the traces of the first wave of the Black Death in narrative, demographic, juridical and environmental sources. It also draws attention to some indicators of the economic condition of Poland in the early second half of the 14th century.

**Chronicles**

Two most important descriptions of the Black Death in Poland are contained in *The Oliwa Chronicle*, recording the history of the Oliwa monastery near Gdańsk between 1186 and 1350, and *The Annals of Jan Długosz*, which is the most important narrative source for the history of late medieval Poland. *The Oliwa Chronicle* is a first-hand account of events witnessed by the prominent figures of the Cistercian monastery. Its compilation began in middle of 13th century. Jan Długosz began writing his monumental work after 1455 and continued until 1480. References to both texts have appeared in many Polish and foreign studies. In the majority of cases, they have been used to provide evidence in favour of the claim that the Black Death was present in Poland in the middle of the 14th century.[22]

However, relevance of *The Oliwa Chronicle* can easily be questioned because it does not, in fact, concern Poland, but Pomerania and Prussia, which at that time were in the hands of the Teutonic Order. The chronicle contains entries concerning the plague in Mediterranean countries and Germany. One of them includes the following commentary: "The aforementioned plague, which has spread over almost all southern countries, oh horror of horrors!, reached our lands as well; in most of Prussia and in Pomerania it has claimed a multitude of men and women, and it continues to claim them still".[23] With regard to this chronicle, first Robert Hoeniger in the 19th century and then Jarosław Wenta in the late 20th century noted that its author copied the main description of the plague from the so-called Avignon Letter, a well-known and widely distributed account of the pestilence path from India to Europe, with special attention given to its outbreak at the papal court in Avignon[24]. References to the Letter can be found in Flemish *Breve Chronicon clerici anonymi*, Austrian *Continuatio Novimontaniensis*, the chronicle of Detmar of Lübeck and Polish *Rocznik Miechowski*.[25] According to Wenta, the authoritative expert on the *Oliwa Chronicle*, the Prussian version of the Avignon Letter with a short mention of the plague in Prussia and Pomerania was incorporated into the original text of the Chronicle at the turn of the 1350s and 1360s. These were the years when the second wave of the Black Death struck, hence the remark that "it continues to claim them still".[26] For this reason, it is hardly possible to assess whether the account of the scale of the Black Death provided by the author of the *Oliwa Chronicle* is reliable as a source of information about the situation in Prussia, and even less so in the case of the Kingdom of Poland in the 1340s. In spite of that, Benedictow makes use of it in his latest work to validate information found in *Liber civitatis* of Braunsberg (today's Braniewo)[27].

*Liber civitatis* is no longer extant and it has rarely been referred to in specialist literature. Benedictow discusses the content of several passages from this document which have survived in the form of an excerpt published in 1864. The text contains information about a rise in death rate after 24 August 1349, first in Elbing (Elbląg), then also in Königsberg, Marienburg (Malbork) and other Prussian towns[28]. Benedictow bases the entire chronology of the spreading of the Black Death in Prussia and Poland on this account. His contention is that Elbing was the centre where the first victims died on 24 August and from which the plague started spreading further. He concludes that the disease must have reached Elbing in the second week of July and by the autumn of 1349 it attacked nearby towns and moved through Chełmno to Masovia. In the following year, the plague was to reach the two largest Prussian towns: Gdańsk and Thorn (Toruń) and then the northern territories of the Kingdom of Poland[29]. It must be noted, though, that this chronology, as logical as it appears, is based not on the actual information derived from the source, but on its relatively liberal interpretation by Benedictow, consistent with the epidemiological theory. It is crucial to notice that the description of the plague in the only surviving fragment of *Liber civitatis* takes the form of a brief narrative featuring a converted Jew, Rumbold, who arrived in Prussia and is described by this source as a spreader of the deadly disease. This aspect of the text is not mentioned by Benedictow although it should be taken into account as a factor influencing the reliability of the whole account. Stories about Jews and converted Jews acting as carriers of the plague and other diseases were common in various descriptions of European epidemics[30]. As we read in *Liber civitatis* of Braunsberg, the first victims of Rumbold were to be the people of Elbing where about 9000 died. According to the source, the Jew carried out his evil mission of deliberately infecting Prussian Christians between 12 April and 16 October 1349. The same source, however, states that the people were dying between 24 August and 25 December. Interestingly, the text does not make any mention of deaths in the two largest towns, Gdańsk and Thorn, nor does it refer to any increases in death rate in Braunsberg itself. The entry for the last day of 1349 contains information about the death of the bishop of Warmia, Herman. Benedictow introduces him as yet another victim of the Black Death even though the source does not state it directly. Herman's death might have been natural equally well because it is known that he had suffered from ill health for a long time. As early as in 1343, the bishop's health condition made him ask the pope to appoint an auxiliary to assist him. The chapter must have also been long aware of Herman's declining health because the day after his death its members met and elected his successor[31]. Some other sources appear to support Benedictow's interpretation of *Liber civitatis*. Among them there is a letter from Lübeck, dated 1350, informing about court trials of Jews in Thorn[32] and a list of town board members in Gdańsk revealing that in 1350, 5 out of 16 mayors, jurors and aldermen had to be replaced possibly due to their deaths[33]. This, however, does not make it any less surprising that our knowledge of the Black Death in medieval Prussia, potentially one of the most cataclysmic events in its history, is based on circumstantial evidence rather than on sources produced by the highly developed chancelleries of the Teutonic state and cities. It must also be added that Benedictow's decision to use the abovementioned sources to make any claims about Polish lands is questionable because, apart from being ambiguous, they refer to various regions of Pomerania, not the Kingdom of Poland.

The key primary source of the history of 14th-century Poland is *The Annals of Jan Długosz*. It contains a brief description of the plague in the entry for 1348 and a reference to this description again in the entry for the year 1349. The former mentions the Black Death and its different variants in Europe; it also adds that the epidemic broke out in January and continued intermittently for seven months. In the 19th century, Max Perlbach noticed that the passage characterizing the plague was copied by Długosz from Guy de Chauliac's *Chirurgia Magna*, describing the Black Death in Avignon[34]. Perlbach's claim that the Polish chronicler copied a fragment of the fifth chapter of one of the most popular late medieval medical textbooks[35] was supported by other scholars[36] and is currently treated as an indisputable fact[37]. The passages in Długosz about the fear of the disease, which prevented parents from taking care of their children (and vice versa) and about accusations against the Jews were also copied by the chronicler from *Chirurgia Magna*[38] and from *The Oliva Chronicle*[39].

The 1349 entry relates specifically to the situation in Poland: "This year brought the plague to Poland too, and as it spread everywhere, many people among the gentry as well as among peasants died. And when no remedy could be found for this long-lasting vexation, and when the plague not only killed many in houses, but also depopulated whole towns and villages, people convinced themselves that all their troubles fell on them as a divine retribution for their crimes and thus they turned to religious practices. So, they flagellated and birched each other, and humiliated themselves with other forms of penance until God showed his mercy towards them and took away the plague and let the acute mortality cease."[40] The editors of *The Annals* emphasize that the quoted passage draws heavily on the same sources as the ones Długosz relied on writing the 1348 entry, that is the work of Guy de Choliaco and *The Oliwa Chronicle*. It is remarked that Długosz „not having any new material, writes commonplaces" that could be used to describe any epidemic anywhere[41]. Długosz's lack of interest in the Black Death becomes apparent after a simple page count, which reveals that sentences relating to the epidemic take only one of 13 pages documenting events from the years 1348-1350. To illustrate how much the chronicler ignored the plague, it is worth mentioning an entry in which he informs about the death of a Cracow cathedral priest. Długosz writes that the priest was murdered on the king's order, which caused God's righteous wrath: „for this odious murder, detested by Him and all the people … God directed His wrath not only at king Casimir, but at his Polish subjects as well" [42]Interestingly, when Długosz names the punishment inflicted by angry God, it is not, as we would expect, the largest epidemic in European history, but Lithuanian military raids on the territory of the Kingdom of Poland[43].

Thus, the problem with Jan Długosz's work, written over 100 years after the first wave of the epidemic is similar to that with *The Oliwa Chronicle* – the fragment relating to the Black Death is a mere compilation and paraphrase of other sources describing the plague elsewhere in Europe. This does not prevent Benedictow from drawing far-reaching conclusions from Długosz's text. Unfortunately, portions of Benedictow's argument are in fact irrelevant because they are based on the false assumption that Jan Długosz was the author of two different texts: *The Annals of Jan Długosz* (known to the author in its English translation) and *The History of Poland* (read by Benedictow in its original Latin version published in 1711) [44]. In fact, Długosz authored one chronicle and the English text known as *The Annals* is an abridged version (selection made by the English editor) of the Polish translation of the original 12-volume Latin text. Therefore, Benedictow's analysis of how the plague was described in two allegedly different texts by Długosz is rather pointless because any differences in length result from the 20[th]-century editing made for the purposes of the English translation. Benedictow's treatment of Długosz's text is also questionable in how he tries to 'amend' the chronicler's chronology of the plague so that it 'fits' his theory. Benedictow claims that it was Długosz who made a mistake placing the plague in 1348 and 1349, while the Prussian sources (whose reliability has already been contested above) indicate that the disease could not have reached Poland earlier than in 1350. Benedictow also attempts to explain why the length and seasonality of the plague presented in *The Annals* does not fit the general theory of the epidemic, which in the light of the abovementioned doubts about Długosz's account's reliability appears quite pointless again.

In order to strengthen his argumentation, Benedictow refers to two additional Polish historical sources. The first one is Marcin Kromer's 16[th]-century work, which the historian characterizes as a „mostly independent account"[45]. However, Kromer's history was written 200 years after the events in which Benedictow is interested. Moreover, it must be remembered that although Kromer is viewed as a quite critical reader of Długosz, *The Annals* "were directly or indirectly his main source of information".[46] The other text referred to by Benedictow in his 2004 book is the 16th-century chronicle of Maciej Miechowita or rather some of its fragments which have been frequently quoted in German historiography, especially by Georg Sticker.[47] Benedictow treats this source as an independent text adding validity to his conclusions drawn from Długosz's work, but as far as the part covering the period before 1480 is concerned, Maciej Miechowita's chronicle is perceived by scholars as a "somewhat revised and supplemented summary of Jan Długosz's history", so it can hardly be trusted as a source of information about the outbreaks of the plague in Poland.[48]

Other historical sources do not explicitly confirm the presence of the first wave of the Black Death on Polish territories either. Among them, researchers draw attention to the *Miechów Annals*, which describe the plague in Hungary under the date 1349, while the epidemic in the Kingdom of Poland appears much later, under the date 1360. Wojciech Drelicharz, an expert on Polish historical annals, claims that the compiler of the *Miechów Annals* mistakenly placed the description of the plague during the reign of Casimir in the entry referring to the year 1360 instead of 1349.[49] Such an arbitrary attitude of Drelicharz to the *Miechów Annals* and their chronology is a result of his conviction that Jan Długosz was right dating the outbreak of the plague in Cracow to 1349 (which, as it has been explained above, is not an unquestionable fact). Moreover, some scholars have also identified similarities between the description of the epidemic in the *Miechów Annals* and the *Avignon Letter*, which further weakens the reliability of the source on the issue of the Black Death.[50] It is worth adding that the 1349 entry in the *Miechów Annals* contains information not only about the plague in Hungary, but also about Polish matters, such as Tartar raids and the king's expedition to Red Ruthenia. The author's silence about the Black Death is a suggestion of its absence, otherwise it would have been included in the entry. Other annals mentioning the plague in the middle of the 14th century are *Spominki Władysławskie*. They include a laconic entry: "Anno domini 1349 pestilencia magna fuit, et homines se affligebant seu flagellabant".[51] This and a few other entries were later copied by the author of *Władysławski Calendar*. It was compiled in Włocławek, the bishop's seat in

Kuiavia, in the northern borderland between the Kingdom of Poland and Teutonic Order's state. The entry did not specify the exact location of the epidemic and did not provide information about where the flagellants were active. It was rather a standard, universal description of the events that could have taken place in any European country at that time.

To conclude, two basic chronicle sources and other sources of lesser prominence mentioning the Black Death in Polish lands copied the descriptions of the plague in Avignon and cannot, therefore, be used to support the thesis of the presence of the first wave of the plague in Poland.

### Demographic sources – Peter's Pence

Studies on the Black Death in Poland have until now completely ignored one of the most important primary sources used in demographic research on the population of late medieval Poland, that is registers of Peter's Pence. It was a tribute to the pope paid since the end of the 10$^{th}$ century, first by rulers and later by the inhabitants of their countries. It took the form of a contribution paid by every house or hearth, collected initially by dukes, then by bishops, and eventually by special papal collectors. Peter's Pence was collected in Poland, England and Scandinavia. Documents produced on this occasion have been used by scholars for over a hundred years in population estimates.[52] They have recently become the basis of Janken Myrdal's assessment of demographic loss in Sweden.[53] Polish Peter's Pence collections have also been described in two monograph studies and a separate publication has been dedicated to the biographies of papal collectors.[54]

After 1318 the method of Peter's Pence payment in Poland changed. The payment for every hearth was replaced by the payment at the rate of 1 denier for each adult person from all social groups (with the exception of non-Christians).[55] It is believed that the reform was politically motivated, as it coincided with Władysław Łokietek's attempts to obtain the pope's approval of his coronation. The king united the key regions under his rule and launched the process of centralization, which put an end to a long period of fragmentation.

Demographic estimates of the size of the population in Poland have been based on the registers of Peter's Pence from the years *after* 1318. Demographers have differed in their attitudes towards research opportunities provided by these sources and towards the methods of population estimation in which these sources are used.[56] A number of characteristics unique to Polish Peter's Pence make these registers a better demographic source material than the Scandinavian or English ones. The most important feature is the fact that Polish Peter's Pence was collected in the form of a poll tax, which enables demographers to make more precise population estimates than in the case of England or Scandinavia where it retained the form of a hearth tax. Moreover, Peter's Pence in late medieval Poland had some political undertones because its collection was supposed to test the effectiveness and efficiency of the bureaucratic machinery of the Kingdom of Poland reviving after decades of decline caused by the country's disintegration. As a result, meticulous documentation of Peter's Pence collection was in the Church's as well as the state's interest. This source has its drawbacks too. One of them is connected with the introduction of the flat rate of Peter's Pence in Polish parishes and dioceses in the second quarter of the 14$^{th}$ century. With time information about the number of individuals contributing to the payment from a given parish or diocese ceased being updated, which undermines the reliability of registers from the demographic point of view. In spite of that, some historians tend to assume that calculations made for the crucial years in the 1340s reflected the actual size of Polish parishes' population.[57] Meanwhile, according to Tadeusz Gromnicki, "In the eyes of the collectors, a taxable entity was no longer an individual member of the parish paying his denier but a parish itself or a vicar at its head. It did not free individual parishioners from paying, as they still paid a denier per head to their vicar, but from the point of view of the collectors it was now the parish that they treated as a taxable entity".[58]

Surviving Polish registers of Peter's Pence cover the period when the Black Death was decimating European nations. It may be assumed that a considerable population loss characteristic of plague time would have left some traces in such documents, in spite of the introduction of the flat rate paid by the whole parish in the 1340s and 1350s. According to Ole J. Benedictow, total mortality rate for all regions of Europe was 60%.[59] Provided that the demographic loss in the Kingdom of Poland was similar, surviving parishioners would have had to pay twice as much in order to maintain the fixed rate of Peter's Pence. There are no traces of such a change in any available sources – throughout the late Middle Ages the rate of Peter's Pence remained at the same level – 1 denier per adult.[60]

At the head of the Church organization in the Polish lands was the archbishop of Gniezno, with several bishops under his authority: the bishops of Poznań, Cracow, Kuiavia (in Włocławek), Płock (in Masovia, at that time independent of Polish kings), Chełmno (in territories conquered by the Teutonic Order) and Wrocław (in Silesia, most of which was at that time independent of Polish kings). For the years 1345-1358 the registers of a papal collector Arnald de Lacaucin have survived. Unfortunately, for most dioceses, all that is available is collective data, sometimes for a few years taken together. That is the case of the bishop of Kuiavia, who paid 1005 *Toruń marks* in 1355 for the previous ten years. The document states that the amount paid for each year between 1346 and 1355 was 100,5 *Toruń marks* (or 62,4 *Cracow marks*).[61] Collective data from other dioceses are more precise and provide information about the exact amounts paid each year (Graph 1). Almost all of them are characterized by relative stability in payments throughout the period of the first wave of the Black Death in Western Europe. One exception is the diocese of Cracow, the second most important diocese in the Kingdom of Poland. Fortunately, surviving registers of Peter's Pence collection from this district take the form of detailed records of payments made by individual parishes.[62] They show that not all sums collected in parishes were eventually sent to Avignon. Hence in Graph 1 the Cracow diocese is represented by two lines. The first one (Cracow max. in Graph 1) informs about the sums collected, whereas the other (Cracow min. in Graph 1) represents the sums sent to the pope. Further in the article, I will refer only to the former.

**Graph 1. Peter's Pence (in marks) paid by Polish dioceses in 1346-1358**

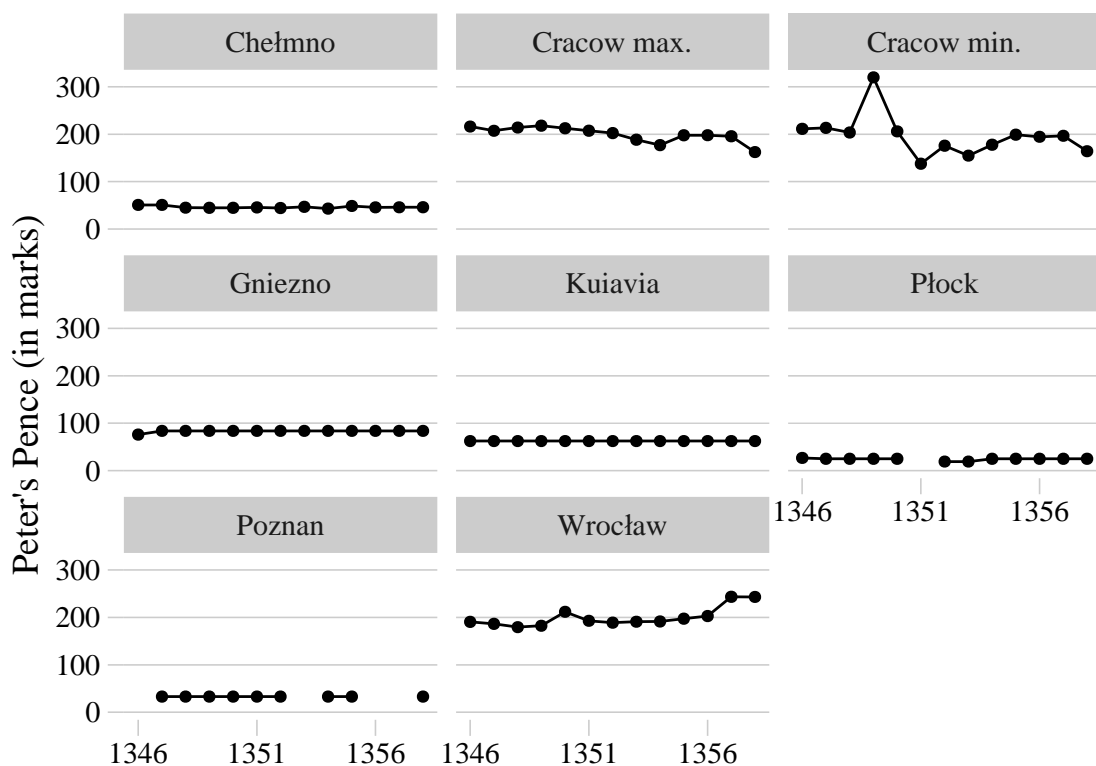

Source of the data: Tadeusz Gromnicki, *Świętopietrze w Polsce* (Kraków: Drukarnia Koziańskiego 1908); *Monumenta Poloniae Vaticana*, vol. 2, ed. Jan Ptaśnik (Kraków 1913), 191-255.

It is assumed in literature that by 1346, shortly before the arrival of the plague in Europe, all parishes in the diocese of Cracow had had the flat rate of Peter's Pence introduced.[63] Between 1346 and 1358 payments from the total number of 536 parishes were registered, though not all of them made regular payments each year. The largest number of paying parishes was registered in 1349 but after that date documents reveal a steady annual decline until 1355, when the number of taxable entities reached the lowest point of fewer than 400 (Graph 2).

**Graph 2. Number of parishes in Cracow diocese paying Peter's Pence in 1346-1358**

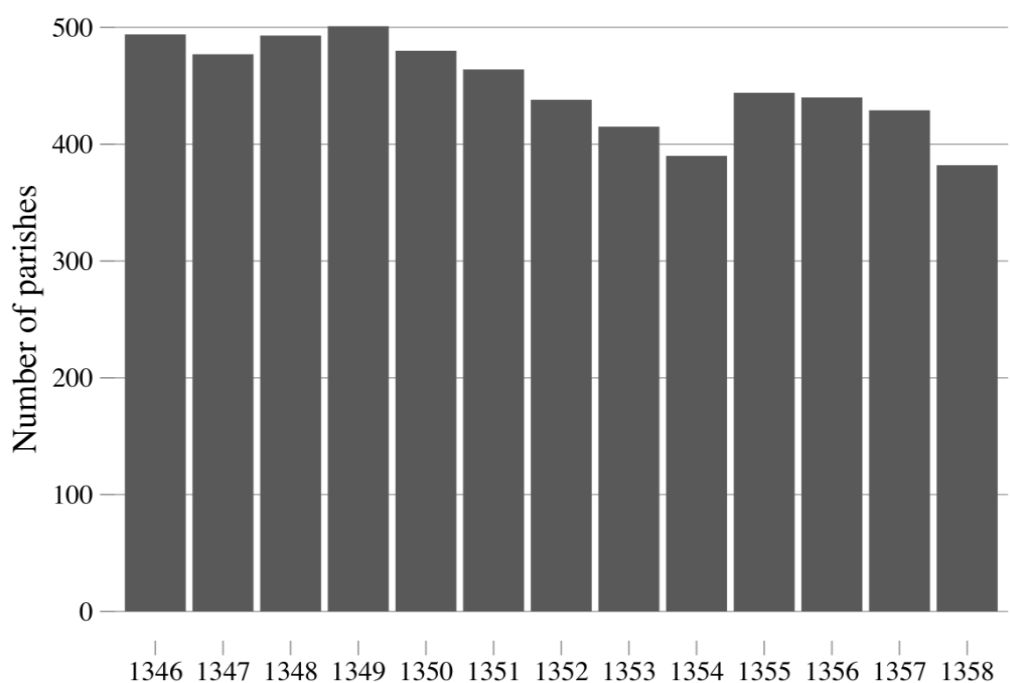

Source of the data: like in Graph 1.

Afterwards, the situation started to improve. The analysis of amounts paid by individual parishes does not show any noticeable changes over the period under study. The vast majority of parishes – 190, paid their Peter's Pence regularly every year; in 136 cases the annual amount did not change throughout the period of thirteen years. The fluctuations, with a periodical decrease and then an increase in payment, concerned 28 parishes, a persistent decrease in payments was recorded in 8 parishes and an increase in 18 parishes. The average amounts (the + character indicates the arithmetic mean) paid by parishes in subsequent years (across the quantiles as well) do not show any significant changes (Graph 3).

**Graph 3. The average amounts paid by parishes in 1346-1358**

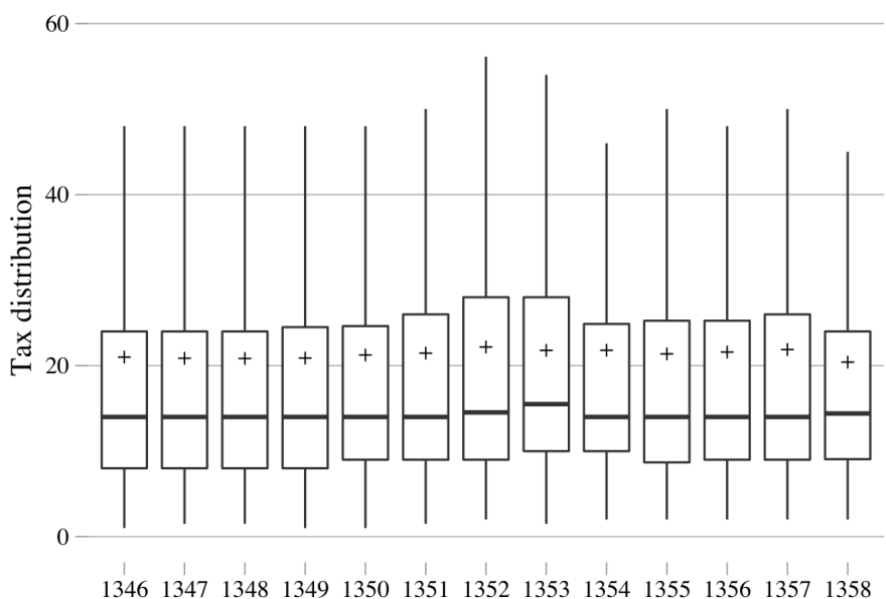

Source of the data: like in Graph 1.

Differences in the amount of tax collected in the whole diocese result from the change in the number of parishes paying to the Holy See in subsequent years (Graph 4).

**Graph 4. Peter's Pence (in marks) paid by Cracow diocese in 1346-1358**

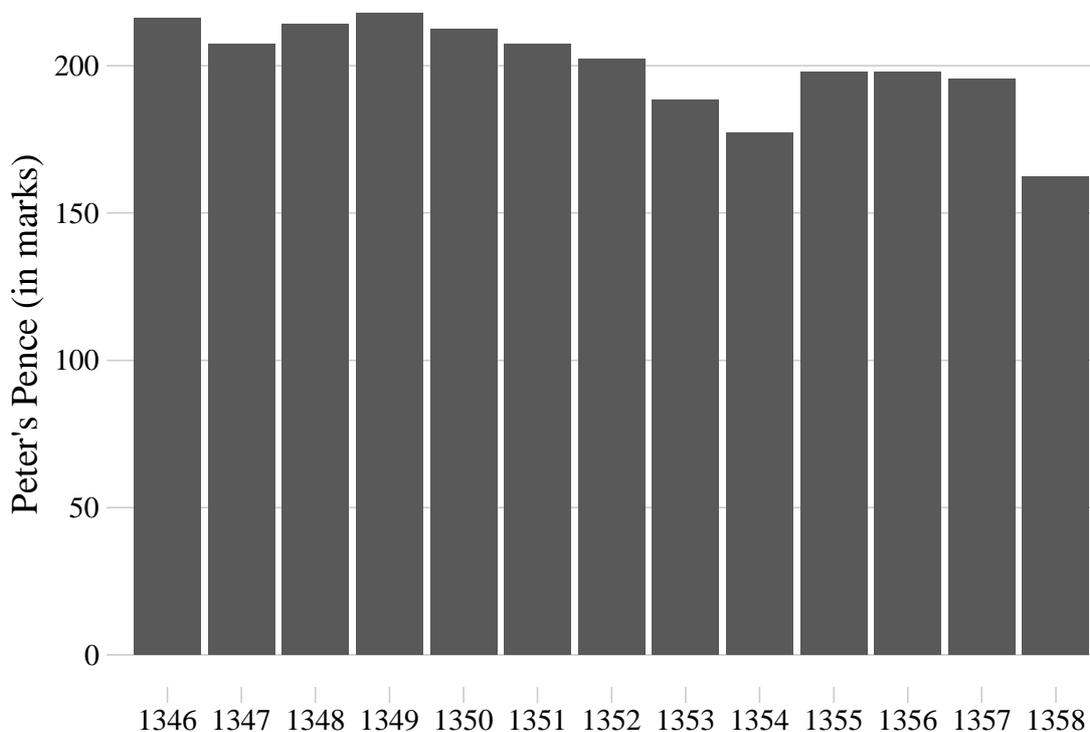

Source of the data: like in Graph 1.

This means that if the Black Death reached the diocese of Cracow, its presence did not manifest itself in the reduction of sums paid by individual parishes, but instead, in the reduction of the number of parishes (Graph 5).

**Graph 5. Changes in the number of parishes and amounts paid in Cracow diocese 1346 – 1358**

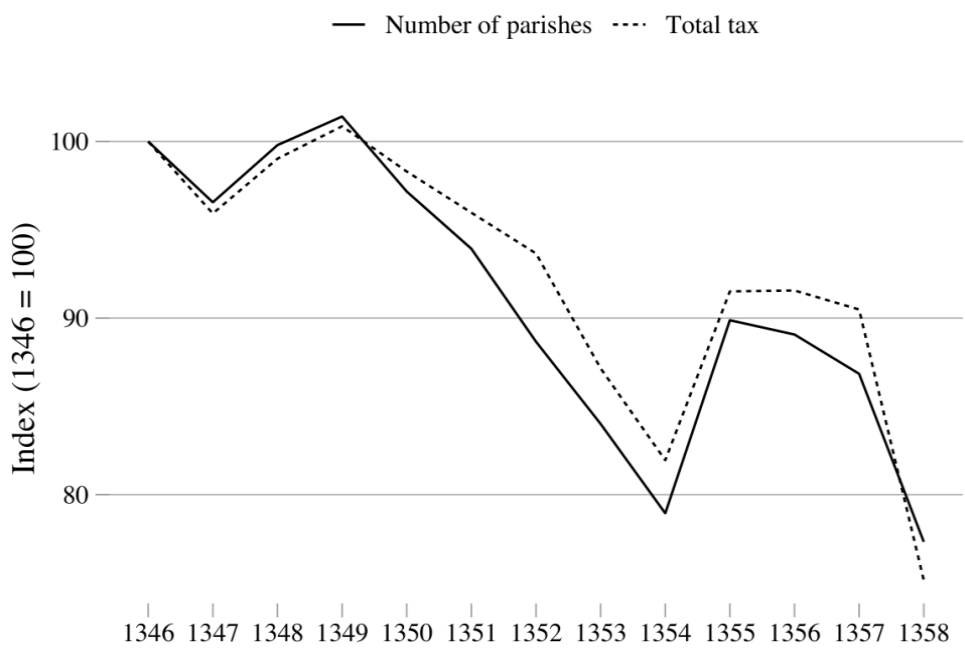

Source of the data: like in Graph 1.

Meanwhile, the spatial analysis of payments indicates that the greatest decrease in the number of paying parishes occurs in the eastern regions of the Cracow diocese, such as the Lublin archdeaconate, the Nowy Sącz deanery and the eastern parishes of the Radom archdeaconate (Map 1). Tadeusz Ładogórski, demographer who successfully used the data on Peter's Pence payments to estimate the population size in Poland in the first half of the 14[th] century, suggested that lack of payments from peripheral parishes should not be explained by the epidemics, but by dishonesty of papal collectors, who embezzled the funds, especially in remote parishes, and never recorded their collection.[64] Additionally, it is possible that the reasons for the disappearance of paying parishes from records ought to be sought also in the country's political circumstances.

**Map. 1.** **Parishes of the Diosece of Cracow in the Kingdom of Poland and Silesia paying Peter's Pence in the middle of the 14th c.**

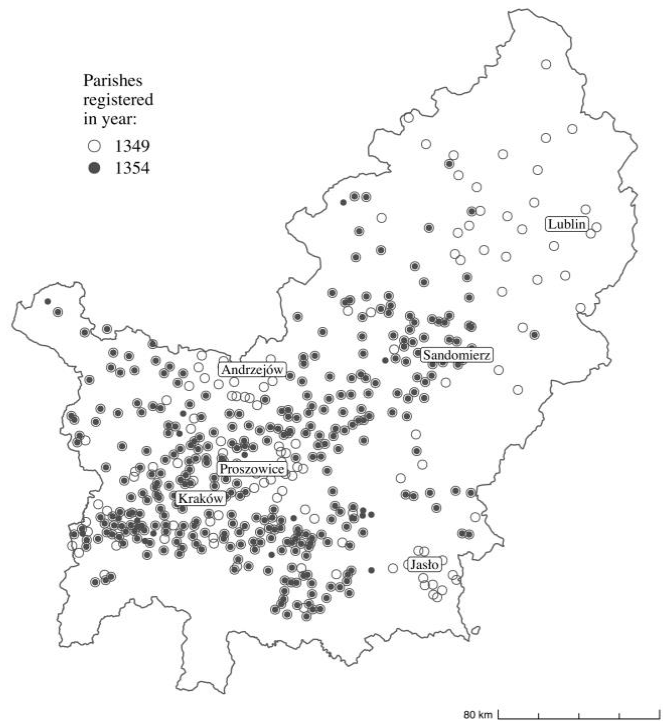

In September 1349, king Casimir began the action of conquering the Ruthenian lands (Wolhynia and Halych Ruthenia), i.e. the territories which were the subject of competition between Poland, Hungary and the Grand Duchy of Lithuania. As an act of retaliation, in May 1350 the Lithuanian army invaded the eastern part of Little Poland (the Sandomierz Land), some territories of central Poland and Masovia, most of which were parts of the diocese of Cracow. This did not discourage the Polish king, who in early 1351 was granted by Pope Clement VI a four-year exemption from half of the tithes, and received support in building the structures of the Latin Church in the Ruthenian lands. Correspondence with the pope did not contain any mention of the plague. In the same year the Lithuanian army invaded eastern Polish lands again, and the following year the Tartars, allied with the Lithuanians, invaded Ruthenia and reached the eastern territories of the diocese of Cracow, plundering the Lublin Land. Despite the truce of 1353, Lithuanians raided Ruthenian and Polish lands again, reaching as far as Zawichost.[65] It is very unlikely that military operations involving Polish, Hungarian, Lithuanian and Tatar forces would have been conducted in conditions of raging plague. It can therefore be assumed that the most plausible reason for a slight the decline in the value of Peter's Pence paid by the diocese of Cracow was the devastation of its eastern territories as a result of the ongoing wars.

It can therefore be concluded that the registers of Peter's Pence payments do not provide grounds for claiming that in the 1340s and 1350s there was a decline in the population of the Kingdom of Poland, unlike Sweden and Norway, where the fall in revenue from this tax was a symptom of a significant population decline due to the epidemic.

**Cracow court records**

The key type of primary source documenting the manifestations of social and economic life in the late Middle Ages is court books kept for particular social groups. In the case of Poland, however, it should be emphasized that the books of noble and peasant courts were not introduced until the turn of the 14$^{th}$ and 15$^{th}$ centuries. The first municipal judicial registers had appeared earlier than that. The majority of surviving urban court books come from the towns which in mid-14$^{th}$ century belonged to the Teutonic Order and became Polish only a hundred years later. One exception is the municipal books of the country's former capital, Cracow, which had been probably kept since the turn of the 13$^{th}$ and 14$^{th}$ centuries. The two oldest preserved books are the bench books (*libri scabinales*) covering the years 1300-1375 and 1365-1376, recording information on transactions of purchase and sale of real estate, as well as on various matters relating to inheritance and credit.[66] They do not, however, record any information about the outbreak of the plague in the city. Quantitative analysis of the recorded court notes shows a significant reduction in the number of registered cases from the mid-1340s to the mid-1360s. A similar analysis of data from the second of the two surviving court books reveals an increase in the number of recorded cases. The exact chronology of the downward trend – between 1345 and 1364 – does not allow to associate it directly with the emergence of the first wave of the epidemic, because it reached Europe at the earliest in the middle of 1347. Similarly, the sudden increase in registrations since 1365 cannot be explained by the rapid population recovery after a possible second wave of the Black Death. The historians of the late medieval Cracow municipal office, analyzing in detail the content of court records, point out that "at least since 1340, two separate bench books were kept in the office, of which one is now lost".[67] In other words, two series of court books functioned simultaneously in the town. The older of the aforementioned surviving books (1300-1375) represents the first series, whereas the other one (1365-1376) comes from the second series, which explains the overlapping of dates and a somewhat different nature of recorded entries. As it is suggested by Bożena Wyrozumska, "It is therefore necessary to accept the fact that the oldest municipal court book contains only a selection of entries, but it is not known how the selection was made".[68] This observation is important in the context of the problem of the Black Death in the middle of the 14th century. We cannot exclude the possibility that the decrease in the number of entries in the surviving city book was compensated by an increase in the frequency of records in the second, lost, one. It is also possible that economic transactions were recorded in the books of the city council instead. The oldest surviving source of this kind comes from 1392, but it is known that the council had kept such books at least since the middle of the 14th century. The falling number of surviving court records since the mid-1340s cannot therefore be interpreted as a symptom of the demographic crisis in the city. Moreover, the preserved records do not mention the plague, and the number of wills contained in the books during the first three quarters of the 14th century did not increase (Graph 6).[69]

**Graph 6. Number of entries and wills in the bench books in Cracow in 1300- 1376**

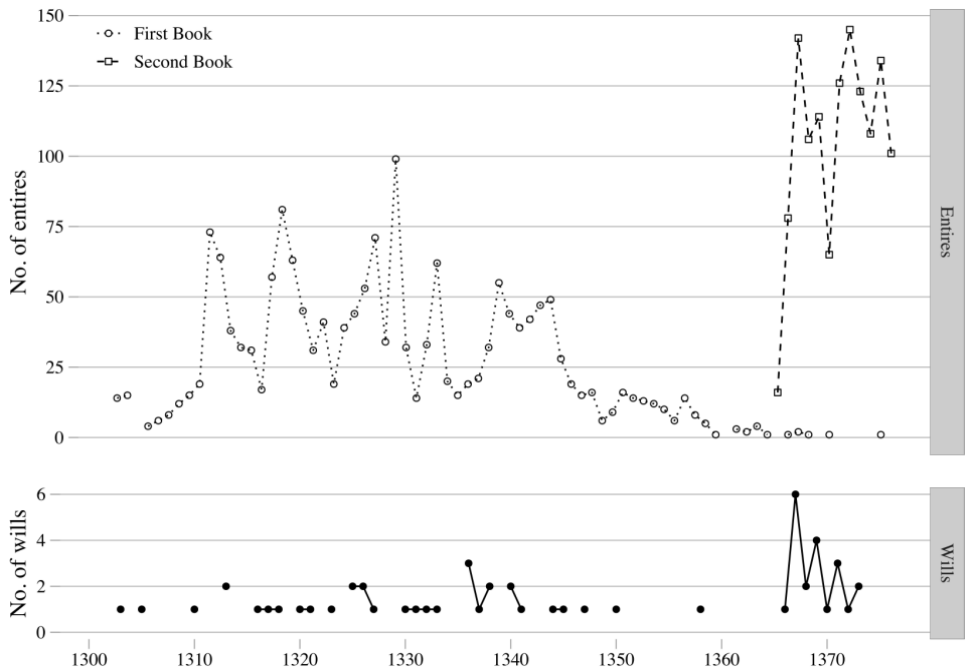

Sources of the data:

*Najstarsze księgi i rachunki miasta Krakowa*, eds. Franciszek Piekosiński and Józef Szujski (Kraków, 1878); *Księgi ławnicze krakowskie 1365-1376 i 1390-1397*, eds. Stanisław Krzyżanowski (Kraków, 1904); Jakub Wysmułek, *Testamenty mieszczan krakowskich (XIV-XV wiek)*,(Warszawa, 2015), 347-364.

**Palynological data**

In recent years extensive databases with data from a large number of pollen sites in Central Europe have been created. Some of them have been used by Adam Izdebski's team in their studies in which palynological data is an alternative "biological measure", which can serve as a useful indicator of human activity.[70] Palynologists extract sediment cores from the lake floor and peatland, establish a chronology of layers of sediments, and count pollen grains of various plant taxa within each layer. In the resulting data set, cereal pollen grains are selected from among all others, because of their significance as anthropogenic indicators, or more specifically as indicators of human economic activity. Changes in pollen structure reflect changes in vegetation in a given area and are indicative of the intensity of people's agricultural activity. It is important to note here, though, that discussions continue among scholars as to the size of area around the site for which the collected pollen data is reliable and representative. Pollen data have also been used to test the scale of epidemics, such as Justinianic plague or even the Black Death.[71]

In the case of Poland, preliminary data, unfortunately of uneven quality, reveal no significant crisis in agricultural production in the years of the Black Death in Western Europe. In Little Poland, for which written sources in the form of Peter's Pence registers are available, palynological data from the period of German colonization and the turn of the Middle Ages show an increase in the proportion of cereal pollen grains in pollen samples. The most notable growth is in the proportion of rye, which was one of the basic breadmaking cereals.

The data for Great Poland show relative stability and then a small drop in agricultural output in the later Middle Ages. The extent of this decrease, however, does not imply a population fall on the scale comparable with that during the Black Death in Western Europe. What is clearly visible in the case of Great Poland though is a significant rise in the proportion of cereal pollen at the beginning of the Early Modern period, which reflects a surge in the production of cereals for export.[72]

For the purposes currently conducted studies new pollen databases will be created, so presented preliminary conclusions will be verified and revised. One tentative conclusion that can already be made is that neither Great-Polish nor Pomeranian data confirm the fall in anthropopressure in the middle of the 14[th] century.[73]

**Demographical data and retrogressive method**

The aforementioned registers of Peter's Pence, containing detailed data for all parishes of the diocese of Cracow and general data for the remaining dioceses, became the basis for the population estimates in three most important regions of the Kingdom of Poland: Little Poland, Great Poland and Masovia. Historians and demographers using these data in their studies applied diverse methods, as a result of which details of their findings differed. They all agreed, however, that information about the payments of Peter's Pence was the only reliable basis for the estimation of the number of people living in Poland in later Middle Ages. The agreed rough estimate for the population in the three regions around 1340 was 1.2 million.[74] A similarly reliable source of information that could be used to estimate the size of Polish population was not produced until the middle of the 16th century.

Relatively well-preserved treasury sources from the second half of the 16th century, in particular from periods when the tax base was updated, can be used for the population estimates in Early Modern Poland. The most valuable category of sources was: tax registers (Pol.: *rejestr poboru*) for the land tax from the years 1578–1581. The tax was paid by all villagers and inhabitants of small towns, regardless of their status; sometimes it was levied on every head of the household, sometimes it was based on the amount of arable land. In demographic literature land tax registers are regarded as very reliable bases for population estimate.[75] Despite the fundamentally different approaches to sources and the use of different converters, historical demographers tend to agree that the estimated population of Little Poland, Great Poland and Masovia around 1580 ranged between 2.89 and 3.15 million.[76]

Estimates of the population of the Polish lands around 1340 compared with the results of studies concerning the year 1580 allowed Irena Gieyszorowa to determine the population dynamics, i.e. the population change resulting from natural growth and migration. In the three main regions of Poland, Gieysztorowa identified a gradual increase in the population over 240 years of 0.38% annually.[77]

**Table 1. Population dynamics in three major regions of the Kingdom of Poland in the 14th-16th centuries.**

| Years | Great Poland | | | Little Poland | | | Masovia | | | Total | | |
|---|---|---|---|---|---|---|---|---|---|---|---|---|
| | Number of people in thousands | Total growth % | Annual growth % | Number of people in thousands | Total growth % | Annual growth % | Number of people in thousands | Total growth % | Annual growth % | Number of people in thousands | Total growth % | Annual growth % |
| **Cir. 1340** | 560 | 120 | 0.33 | 440 | 180 | 0.43 | 250 | 156 | 0.39 | 1250 | 148 | 0.38 |
| **Cir. 1580** | 1230 | | | 1230 | | | 640 | | | 3100 | | |

Source of data: Irena Gieysztorowa, "Ludność", in Antoni Mączak (ed.) *Encyklopedia historii gospodarczej Polski do 1945 roku,* vol. 2 (Warszawa, 1981), 431.

In comparison with other countries, population dynamics in Poland in the later Middle Ages was relatively high, while in the Early Modern period it could be assessed as average. It is estimated that the population growth of the entire continent since the mid-15th century ranged between 0.25 and 0.28% annually, with significant regional differences. The figures for the most developed European countries are considerably higher than the Continental average. The growth rate in England for most of the 16th century was well above 0.5%, reaching as much as 0.9% in some decades[78]. For the German countries it was 0.7% between 1520 and 1570, similarly to Northern Netherlands (United Provinces) and Spain (0.6%). The figures for Southern Netherlands were lower than for Poland (0.21%).[79] In her analysis, Gieysztorowa had access to data from only two years – 1340 and 1580. She made an assumption that neither in the late Middle Ages nor at the beginning of the Early Modern period were Polish lands affected by any significant demographic catastrophe. Provided the Black Death had reached Poland and half of the population had died, natural growth after 1352 would have had to exceed 0.6% in order to have a population of 3 million in 1580, which is unrealistic. Birabean estimated that immediately before the Black Death, at the time of prosperity, European population growth was at 0.3% per year, falling to 0.11% after the crisis.[80] With a 0.67% rate or bigger, Polish demographic explosion in the turn of Middle Ages would have been an inexplicable exception in Europe. Even when we take migration into account, it does not seem possible that they could have accounted for such a considerable increase in population. Benedykt Zientara, a specialist in late medieval colonisation, estimated the number of settlers who came to Poland from the West in the 12th to 15th centuries at 100 000.[81] It was not a small group, but certainly not a decisive factor in the almost threefold increase in population over 240 years.

**Institutional and economic context**

Ole J. Benedictow supports his thesis of the plague present in the Kingdom of Poland with an economic argument about the fluctuation of prices and wages. In his work published in 2004, he quotes William Abel's work on the crisis of the later Middle Ages and the author's observation that changes in prices and wages in late medieval Cracow follow the same pattern as in England or France, where the Black Death did occur. In all three cases the prices of grain decreased while wages grew.[82] In Benedictow's new work on the Black Death published in 2021, the author includes a graph presenting grain prices and wages derived from William Abel's book[83]. Abel's database included Cracow prices and wages of town officials, carpenters, and masons. It had originally been compiled in the 1930s by Julian Pelc[84]. However, the quality of data for the late Middle Ages is questionable because of the fragmentary and random nature of surviving source material. For a long period between 1350 and 1525 (the time frame proposed by Abel and then echoed by Benedictow), that is 175 years, wheat prices are known from only 15 years, rye prices from 24 years, barley prices from 9 years, oats prices from 44 years (Graph 7). The wages of town scribes are known from 18 years, carpenters' wages are known from 3 years, while the wages of masons over this period are not documented in available sources at all[85]. Therefore, the data used by Abel and Benedictow are not sufficient to allow to form conclusions about the economic condition of the Kingdom of Poland (more on this topic below), not to mention any conclusions concerning population trends. In the context of this article, it is also important to emphasize that almost all the fragmentary data mentioned above come from the turn of the 14th and 15th centuries and later, long after the first wave of the Black Death ocurred in Europe.

Graph 7. Number of yearly prices in Cracow by decades of Late Middle Ages.

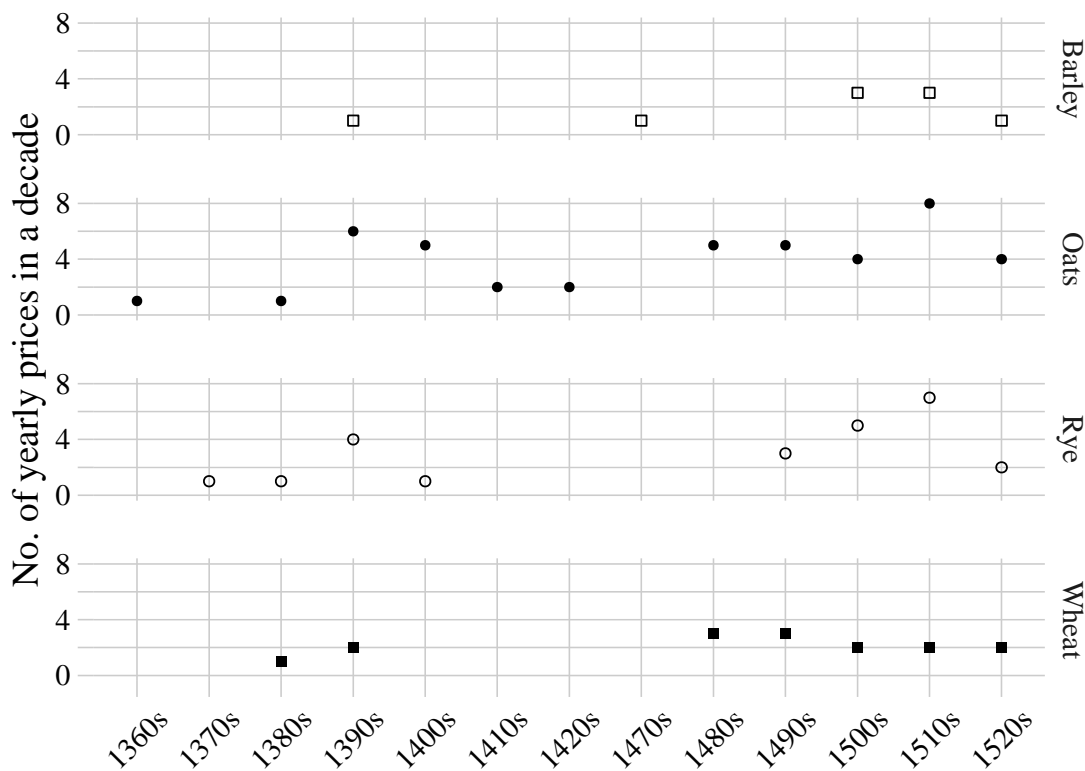

Source of data: Julian Pelc, *Ceny w Krakowie w latach 1369-1600* (Lwów, 1935), 6-15.

Although our sources for economic history are generally not as reliable as in Western Europe, it can be amply demonstrated on their basis that 14th-century Kingdom of Poland enjoyed the period of economic growth uninterrupted by the epidemic. One may have reservations about the image of King Casimir the Great (1333-1370), idealized by Polish historiography as early as in the 15th century, but his achievements in unifying the country after the period of feudal fragmentation are undeniable. He followed the rule: *Unus princeps, unus ius, una moneta in toto regno.*[86] His successes are often summarized with a saying "Casimir the Great inherited Poland made of wood and left it made of stone", a paraphrase of Jan Długosz's famous passage in his late medieval chronicle. Attempts to trace any radical change in king's behaviour or policies that could be indicative of the sudden outbreak of the epidemic do not bring any results. Reconstructions of Casimir's itineraries reveal that between 1347 and 1351 the king travelled extensively over Little and Great Poland and to neighbouring Masovia. Beginning in 1351, for a few years, political unrest and military operations kept the king mostly in Little Poland and Red Ruthenia.[87] Documents issued by the royal office do not reveal any decline in the king's socio-political activity or do not imply that the king avoided residing in the capital city (which, as the chroniclers claimed, was struck by the plague) either in late spring or summer months when the plague would have spread most rapidly. Although the king entreated the pope to grant special plenary indulgences at the moment of death (*absolution plenaria, in articulo mortis*) in 1347, 1348 and 1350, it must be noted that he had sent similar requests to the Holy See in the 1330s and repeated them later in the 1360s, which may suggest that they were not a direct response to the imminent threat posed by the actual outbreak of the plague, but by the awareness of its existence elsewhere.[88]

Also the ruler's investment activities did not appear to be in decline in the second half of the 14th century. Of the 23 royal initiatives to surround the largest cities with defensive walls, most started after 1350.[89] Moreover, the king "implemented a large scale policy of building territorially compact property units with the city as an economic centre and often a castle as a local centre of power".[90] The expansion of the royal domain, especially in Little Poland and in Red Ruthenia, was achieved by means of colonization.[91] Soon the greatest noblemen followed suit. During the first twenty years of Casimir's reign (1333-1350) 87 cities and towns were established in Little and Great Poland; during the last two decades (1351-1370), 132 more were created.[92] Simultaneously, in years 1333-1350 in 53 villages in Great Poland the German law was introduced and they were reorganized accordingly. Between 1351 and 1370 similar changes took place in further 95 Great Polish villages.[93] The process of introducing the German law can be traced in other regions of the kingdom as well.[94] The general growth in settlement is visible in the number of new villages recorded in primary sources. In Great Poland, which has been studied more than other parts of the kingdom, the second half of the 14th century was characterized by the biggest increase in the number of newly registered settlements in the whole Middle Ages.[95] Such a growth would not have been possible, had there been a plague-related demographic and ensuing economic crisis.

The lack of the economic crisis can be further evidenced by information derived from the records of financial transfers from the Kingdom of Poland to Avignon. They included the aforementioned Peter's Pence as well as annates and the papal tithe.[96] The latter was levied on the clergy relatively seldom and was not universal, i.e. the clergy in some countries were exempt from payment. For example, when the papal tithe was levied in 1343 with the aim of raising funds to build galleys to be sent against the Turks, it was not paid by the English, Spanish or the French clergy. It was paid by the Polish clergy though, as was the tithe levied in 1351 and 1355. Half of the sums collected in the 1350s was left at the disposal of the Polish king, who fought against the Tartars and the Lithuanians.[97]

The level of financial contributions from Poland to Avignon, although not particularly significant in the total budget of the papal curia, continued to rise after 1350.[98] It is hard to disagree with Marian Małowist's opinion on Central Europe: "It should be stressed with the utmost conviction that in the second half of the 14th century there was no economic slowdown here. In the kingdoms of Poland, Bohemia and Hungary, there was even an acceleration of economic development".[99] These words were originally written almost half a century ago, and despite the development of research on the extent of the first wave of the Black Death in the countries concerned, no breakthrough research can be said to have challenged this opinion thus far.[100] It also appears that even the better documented cases of successive epidemic outbreaks at the end of the 14th century do not allow us to conclude that the scale of depopulation was comparable to that of Western Europe.[101]

**Conclusion**

The data presented above confirm that in the middle of the 14th century the Kingdom of Poland did not experience depopulation. The analyses of Polish narrative sources from the late Middle Ages, registers of Peter's Pence and Cracow court books, complemented by the analysis of environmental archives and a retrogressive overview of Early Modern demographic data raise serious doubts about the possibility that the Black Death reached the Kingdom of Poland in the middle of the 14th century. None of these sources suggests a rapid and massive (up to 50%) reduction in the population of the main regions of the Kingdom of Poland, which challenges the existing theories about the pace and pattern of the spreading of the epidemic in Europe. The Polish lands in the middle of the 14th century were not, after all, an isolated area, and the king and Polish cities maintained trade and diplomatic contacts with all major European centers. Perhaps a critical analysis of Central European chronicles and yearbooks should be undertaken once again, and a wider and better use of natural archives would be worthwhile.[102] Only an interdisciplinary approach will allow us to answer the question why the Black Death did not reach the Kingdom of Poland in the middle of the 14th century, and even if it did, why its presence was not recorded in historical mass sources and is not reflected in the environmental material. Perhaps the discussion on the geography and scale of the epidemic should take greater account of the environmental specificity of the region, the seasonal nature of the epidemic - the time of its arrival in a given country, and finally the physical condition of the inhabitants, especially in regions that were previously severely affected by the Great Famine during the second decade of the 14th century.

**Acknowledgements**

This work was supported by the National Programme for the Development of Humanities of Polish Ministry of Science and Higher Education during the years 2016-2020.

---

[i] Philip Ziegler, *The Black Death* (London, 1969); William M. Bowsky (ed.), *The Black Death. A Turning Point in History?* (New York 1971); John Hatcher, *Plague, Population and the English Economy 1348-1530* (London 1977); David Herlihy, *The Black Death and the Transformation of the West*, ed. S. K. Cohn (Cambridge -Mass., 1997).

[2] Stephanie Haensh et ali., "Distinct Clones of Yersinia Pestis Caused the Black Death," *PLoS Pathogens* VI (2010), e100134; Bos Kirsten et ali., "A Draft Genome of Yersinia Pestis From Victims of the Black Death," *Nature* CCCCLXXVIII (2011), 506-10; Cui Yujun et ali., "Historical Variations in Mutation Rate in an Epidemic Pathogen, Yersinia Pestis," *Proceedings of National Academy of Sciences* CX (2013), 577-582; Maria. A. Spyrou et ali., "Historical Yersinia Pestis Genomes Reveal the European Black Death as the Source of Ancient and Modern Plague Pandemics," *Cell Host & Microbe* XIX (2016), 874-881; Deneb Cesana et ali, "The Origin and Early Spread of the Black Death in Italy: First Evidence of Plague Victims from 14th-century Liguria (Northern Italy), "*Anthrolopogical Science* CXXV (2017), 15-24; Monica H. Green, "Putting Africa on the Black Death Map: Narratives from Genetics and History," *Afriques* IX (2018),1-46; Monica H. Green, "The Four Black Deaths", *American Historical Review* CV, (2020), 1601-31; John Aberth, The Black Death. A New History of the Great Mortality in Europe, 1347-1500 (New York-Oxford, 2021), 1-13; Philip Slavin, "Out of the West: Formation of a Permanent Plague Reservoir in South-Central Germany

(1349-1356)", Past and Present CCLI (2021), 3-51; Ole J. Benedictow, The Complete History of the Black Death (Woodbridge, 2021), 3-57.

[3] Richard C. Hoffmann, *An Environmental History of Medieval Europe* (Cambridge, 2014), 290; Joris Roosen and Daniel R. Curtis, "The 'Light Touch' of the Black Death in the Southern Netherlands: an Urban Trick?," *The Economic History Review* LXXII (2019), 32-56.

[4] Ole J. Benedictow, *The Black Death 1346-1353* (Woodbridge, 2004), 380-84.

[5] Sevket Pamuk, "The Black Death and the Origins of the Great Divergence across Europe, 1300-1600," *European Review of Economic History* XI (2007), 289-317; Paolo Malanima, "The Economic Consequences of the Black Death," in Elio Lo Cascio (ed.) *L'impatto della "Peste Antonina"* (Bari, 2012), 311-328; Alexandra M. De Pleijt and Jan Luiten van Zanden, "Accounting for the "little Divergence": What Drove Economic Growth in Pre-industrial Europe, 1300-1800?," *European Review of Economic History* XX (2016), 387-410.

[6] Robert Hoeniger, *Der Schwarze Tod in Deutschland* (Berlin, 1882), 35-38; Georg Sticker, *Abhandlungen aus der Seuchengeschichte und Seuchenlehre,* vol 1*: Die Pest* (Giessen, 1908), 67-68; A similar view was presented by Jean Noël Biraben, *Les hommes et la peste en France et dans les pays européens et méditerranés*, vol. 1 (Paris, 1975), 74-85.

[7] Piotr Rutkowski, "Czarna Śmierć w Polsce w połowie XIV wieku", *Studia i Materiały z dziejów Nauki Polskiej* XXVI (1975), 3-31; Jan Tyszkiewicz, *Człowiek w środowisku geograficznym Polski średniowiecznej* (Warszawa, 1981), 122-123; Idem, *Ludzie i przyroda w Polsce średniowiecznej* (Warszawa,1983), 153-154*;* Jerzy Jankowski, *Epidemiologia historyczna polskiego średniowiecza* (Kraków, 1990); Grażyna Klimecka, "Chory i choroba", in Bronisław Geremek (ed.), *Kultura Polski średniowiecznej XIV-XV w*. (Warszawa, 1997), 62; Szymon Wrzesiński, *Epidemie w dawnej Polsce* (Zakrzewo, 2011).

[8] Jerzy Wyrozumski, *Dzieje Polski piastowskiej,* (Kraków, 1999), 274; Stanisław Szczur, *Historia Polski. Średniowiecze* (Kraków, 2002), 423; Marian Małowist, *Wschód a Zachód Europy w XIII-XVI wieku. Konfrontacje sttruktur społeczno-ekonomicznych* (2nd ed.,Warszawa, 2006), 196.

[9] Beata Możejko, "Zarazy w średniowiecznym Gdańsku", in E. Kizik (ed.), *Dżuma, ospa, cholera, W trzechsetną rocznicę wielkiej epidemii w Gdańsku i na ziemiach Rzeczypospolitej w latach 1708-1711* (Gdańsk, 2012), 43-61.

[10] Wojciech Mrozowicz, "Dolny Śląsk w latach 1327-1526", in Wojciech Wrzesiński (ed.), *Dolny Śląsk. Monografia historyczna* (Wrocław, 2006), 130.

[11] Henryk Samsonowicz, *Historia Polski do roku 1795* (4th ed., Warszawa, 1985), 82.

[12] M. Dygo, "Was There an Economic Crisis in the Late Medieval Poland?," *Vierteljahrschrift für Sozial-und Wirtschaftsgeschichte* LXXVII (1990), 305-322; Jerzy Wyrozumski, "Czy późnośredniowieczny kryzys feudalizmu dotknął Polskę?,"in Tomasz Jasiński, Tomasz Jurek, Jan Maria Piskorski (eds.), *Homines et societas. Czasy Piastów i Jagiellonów* (Poznań, 1997), 103-113; Kszysztof Mikulski, "Folwarki zimany koniunktury gospodarczej w Polsce XIV-XVII wieku, " Klio IV (2003), 25-39; Marian Dygo, "Wschód i Zachód. Gospodarka Europy w XIV -XV wieku," in Sławomir Gawlas (ed.), *Ziemie polskie wobec Zachodu. Studia nad rozwojem średniowiecznej Europy*, (Warszawa 2006), 117-194; Piotr Guzowski, "Kryzys gospodarczy późnego średniowiecza czy kryzys historiografii*?*," *Roczniki Dziejów Społecznych i Gospodarczych* LXVIII (2008), 173-93.

[13] Élisabeth Carpentier, "Autour de la peste noire: famines et épidémies dans l'histoire du XIVe siècle," *Annales ESC* XXVII (1962),1062-92.

[14] Robert S. Gotfried, *The Black Death. Natural and Human Disaster in Medieval Europe* (New York, 1983), 68.

[15] Benedictow, *The Complete History*, 577-584.

[16] Benedictow, *The Black Death*, 218-221.

[17] David C. Mengel, "A Plague on Bohemia? Mapping the Black Death," *Past and Present* CXI (2011), 3-34.

[18] Benedictow, *The Complete History*, 591-603.

[19] Agata Chilińska, Urszula Zawadzka, Arkadiusz Sołtysiak, "Pandemia dżumy w latach 1348-1379 na terenach Królestwa Polskiego: model epidemiologiczny i źródła historyczne," in *Funeralia Lednickie*, vol. 10: *Epidemie, klęski, wojny*, (Poznań, 2008),165-174.

[20] George Christiakos et ali., *Interdisciplinary Public Health Reasoning and Epidemic Modelling: The Case of Black Death* (Berlin, 2005); Janken Myrdal, "The Forgotten Plague: The Black Death in Sweden," in Pekka Hämäläinen (ed.), *When Disease Makes History. Epidemics and Great Historical Turning Points* (Helsinki, 2006), 156, 178; Manfred Vasold, "The Diffusion of the Black Death 1348-50 in Central Europe," in Lars Bisgaard, Leif Søndergaard (eds.) *Living with Black Death,* (Odense, 2009), 54; Bruce S. Campbell, *The Great Transition. Climate, Disease and Society in the Late-Medieval World* (Cambridge, 2016), 303.

[21] Marek Cetwiński, "Fuit maxima pestilencia quais in omnibus terris: klęski elementarne w kronikach i rocznikach polskiego średniowiecza," in Tomasz Głowiński and Elżbieta Kościk (eds.), *Od powietrza, głodu, ognia i wojny… Klęski elementarne na przestrzeni wieków,* (Wrocław, 2013): 11-28; Janusz Skodlarski, "Klęski elementarne na podstawie Roczników Jana Długosza," in Ibidem, 29-40.

[22] *Die ältere Chronik von Oliva* ed. Theodor Hirsh, Scriptores Rerum Prussicarum, vol. 1, (Leipzig, 1861), 668-726; *Chronica Olivensis auctore Stanislao, abbate olivensi*, ed. Wojciech Kętrzyński Monumenta Poloniae Historica, vol. 6 (Kraków, 1893), 310-350; Jarosław Wenta, "Dziejopisarstwo w klasztorze cysterskim w Oliwie na tle porównawczym," *Studia Gdańskie* VII (1990), 109-130; *Ioannis Dlugossi Annales seu Cronicae incliti Regni Poloniae*, liber 9, (Varsoviae, 1978) 252, 257.

[23] *Chronica Olivensis auctore Stanislao*, 347.

[24] Hoeniger, *Der Schwarze Tod in Deutschland*, 4, footnote 1; Jarosław Wenta, "List Awinioński w dziejopisarstwie pomorskim i pruskim połowy XIV wieku," *Przegląd Historyczny* LXXIII (1982), 275-281; Andreas Welkenhuysen, "La peste en Avignon (1348), décrite par un témoin oculaire, Luis Sanctus de Beringen" (Édition critique, traduction, élements de commentaire), in *Pascua Mediaevalia. Studies voor Prof. dr. J.M de Smet* (Leuven, 1983), 452-492;

[25] *Breve chronicon clerici*, Recueil des chroniques de Flandre, vol. 3 (Buxelles, 1856), 5-30. *Continuatio Novimontaniesis a 1329-1346*, ed. W. Wattenbach, Monumenta Germaniae Historica, vol. IX, (Hannover, 1851), 669-677; *Die Chroniken der niedersächsischen Städte. Lübeck*, ed. W. Koppman, Die Chroniken der deutschen Städte vom 14. Bis ins 16. Jahrundert, vol. XIX (Leipzig, 1884), 505; Zofia Budkowa, "Rocznik miechowski," *Studia Źródłoznawcze* V (1960), 119-135; Wenta, *Dziejopisarstwo*, 95-97.

[26] Wenta, *List* , 280; Wenta, Dziejopisarstwo, 111.

[27] Benedictow, *The Complete History*, 545.

[28] *Codex diplomaticus Warmiensis*, vol. 2, eds. Peter Woelky, Johann M. Saage (Mainz, 1864), 152.

[29] Benedictow, *The Complete History*, 590.

[30] Cordelia Hess, „Jews and the Black Death in Fourteenth-Century Prussia: A search for Traces in: Fear and Loathing in the North. Jews and Muslims in Medieval Scandinavia and the Baltic region, eds. Cordelia Hess, Jonathan Adams, Berlin 2015, 109-126. See also: Cohn, S. K. (2007). *The Black Death and the Burning of Jews. Past & Present, 196(1), 3–36.*

[31] Stanisław Achremczyk, Roman Marchwiński, Jerzy Przesmycki, *Poczet biskupów warmińskich*, (Olsztyn 1994), 31-34;

Teresa Borawska, Marcin Borzyszkowski, Herman z Pragi (Hermanus de Praga), in Słownik Biograficzny Kapituły Warmińskiej, (Olsztyn, 1996) 88-89.

[32] Urkundenbuch der Stadt Lübeck, vol. 3, (Lübeck, 1871), 105; Hess, Jews, 118.

[33] Joachim Zdrenka, *Urzędnicy miejscy Gdańska w latach 1342-1792 i 1807-1814. Spisy*, vol. 1(Gdańsk 2008), 20-21; Możejko, *Zarazy*, 50.

[34] Max Perlbach, "Über die Ergebnisse der Lamberger Hanschrift für die ältere Chronik von Oliva", *Altpreussische Monatsschrift* 9 (1872), 32.

[35] Guigonis de Caulhiaco (Guy de Chauliac), *Inventarium Sive Chirurgia Magna*, ed. Michael R. McVaugh and Margarete S. Ogden, vol. 1 (Leiden, 1997), 116-121.

[36] Heinrich Zeissberg, *Die Polnische Geschichttschreibung des Mittelalters*, (Leipzig, 1873), 304; Władysław Semkowicz, *Rozbiór krytyczny Dziejów polskich Jana Długosza (do roku 1384),* (Kraków, 1887), 365.

[37] Original text of Guy de Chauliac: Et hoc manifeste vidimus in illa ingentii et inaudita mortalitate que apparuit nobis in Avinione anno domini millessimo ccc xlviii, pontificatus domini Clementi sexti anno sexto, in in servicio cuius sui gracia licet indignus tunc existebam. Et non displicaet quia propter ipsius mirabilitatem et previdenciam si iterum accideret narrabo eam. Incepit autem predicta mortalitas nobis in mense Ianuarii, et duravit per septem menses. Eta habuit duos modos. Primus per duis menses cum febre continua et sputo sanguinis, et isti moriebantur ifra tres dies. Secundus fuit per residuum temporis, cum febre eciam continua et apostematibus et antracibus in exterioribus, potissime in subasellis et inguinibus, et moriebantur infra quinque dies. Et fuit tante contagiositatis, specialiter que fuit cum sputo sanguinis, quod non solum morando sed eciam inspiciendo unus reciebat de alio, in tantum quod gentes moriebantur sine servitoribus et sapeliebantur sine sacerdotibus; pater non visitabat filium, neque filius patrem. Caritas erat mortua, spes prostrata. Et nomino eam ingentem quia totum mundum vel quasi occupavit. Inceperat autem in oriente, et ita, sagittando mundum, pertransivit per nos versus occidentem. Et fuit ita magna quod vix quartam partem gencium dimisit (Guy de Chauliac, *Inventarium* s. 117-118) and fragment of Jan Długosz's chronicle: Incepit autem pestis predicta epidemie in mense Ianuario in pontificatu Clementis sexti anno eius sexto et duravit per septem menses continuos, duobus modis agitata. Primus quidem tendebatur duobus mensibus per febrem continuam et sputum sanguinis moriebanturque pacientes infra triduum. Secundus tendebatur quinque mensibus per febrem similiter continuam, apostemata et anthraces, que in exterioribus potissime sub ascellis et inguinibus erumpebant, qui pacientes infra dies quinque extinguebat. Tante uterque contagionis, ut non ex conversacione tantummodo et hanhelitu, sed ex solo aspectu infeccio pestifera sequertur. Horrebantque et fugiebant parentes curare natos et nati parentes, execrate alii alteros, videbaturque caritas mortua, spes prostrata. Ab oriente in occidentem se protendens, totum fere mundum adeo infecione sua attriverat, ut vix quarta pars mundi hominum putaretur relicta.*(Ioannis Dlugossi Annales* s. 252.)

[38] Guy de Chauliac, *Inventarium* s. 118

[39] See the comments of editors of Polish translation: Jan Długosz, *Roczniki czyli Kroniki sławnego Królestwa Polskiego*, book 9 (1300-1370), eds. Jan Garbacik and Krystyna Pieradzka, 2nd ed., (Warszawa, 2009), 319 note:11 and 12.

[40] *Ioannis Dlugossi Annales*, 257. The English translation of the cited passage comes from: *The Annals of Jan Długosz. An English Abridgement by Maurice Michael, with commentary by Paul Smith* (Chichester, 1997), 298-301.

[41] Jan Długosz, *Roczniki,* 325, note 26.

[42] *Ioannis Dlugossi Annales*, 258.

[43] *Ioannis Dlugossi Annales*, 258-259.

[44] Benedictow, The complete history, 588.

[45] Benedictow, The complete history, 593.

[46] Henryk Barycz, *Szlakami dziejopisarstwa staropolskiego. Studia nad historiografią w XVI- XVIII w*. (Wrocław, 1981), 86.

[47] Sticker, *Abhandlungen aus der Seuchengeschichte,* 67-68.

[48] Antoni Borzemski, *Kronika Miechowity. Rozbiór krytyczny* (Kraków, 1890), 6-7; Ludmiła Krakowiecka, *Maciej z Miechowa. Lekarz i uczony Odrodzenia* (Warszawa, 1956), 199; Wojciech Drelicharz, *Unifying the Kingdom of Poland in Mediaeval Historiographic Thought* (Cracow, 2019) , 418.

[49] *Annales Miechovienses*, ed. August Bielowski, Monumenta Poloniae Historica, vol. 2 (Lwów1872), 885, 886; Budkowa, *Rocznik Miechowski*, 125-126; Wojciech Drelicharz, *Annalistyka Małopolska XIII-XV wieku: kierunki rozwoju wielkich roczników kompilowanych* (Kraków, 2003), 282.

[50] Wenta, *List*, 280.

[51] *Kalendarz i Spominki Włocławskie*, ed. Brygida Kürbis, Monumenta Poloniae Historica, 2 ser, vol.6 (Warszawa, 1962), 84.

[52] Johan E. Sars, "Til Oplysning om Folkem ængdens Bevægelse i Norge Fra det 13.\ Til det 17. Aarhundrede," *Historisk Tidsskrift* VII (1882), 281-387; Felix Liebermann, "The Peter Pence and the Population of England about 1164," *The English Historical Review* XI (1896), 744-747; Yngve Brilioth, *Den påfliga beskattningen af Sverige : intill den stora schismen* (Uppsala 1915); Asgaut Steinnes, "Roma-skatt og folketal," *Historisk tidsskrift* (Norwegian) XXXII (1940-42), 137-182; Kåre Lunden, "Four Methods of Estimating the Population of Norwegian District on the Eve of the Black Death," *Scandinavian Economic History Review* XVI (1968), 1-18.

[53] Myrdal, *The Forgotten Plague*,141- 186; Myrdal, "The Black Death in the North: 1349-1350," in *Living with Black Death,* 63-84.

[54] Tadeusz Gromnicki, *Świętopietrze w Polsce* (Kraków, 1908); Jan Ptaśnik, "Kolektorzy kamery apostolskiej w Polsce Piastowskiej," in *Rozprawy Akademii Umiejętności*, *Wydział Historyczno-Filozoficzny*, vol. 1 (Kraków, 1907), 1-79; Erich Maschke, *Der Peterspfenning in Polen und den deutschen Osten* (Leipzig, 1933).

[55] Gromnicki, *Świętopietrze*, 26; Tadeusz Ładogórski, *Studia nad zaludnieniem Polski w XIV wieku* (Wrocław, 1958), 71-74.

[56] Tadeusz Ladeberger, *Zaludnienie Polski na początku panowania Kazimierza Wielkiego* (Lwów, 1930); Ładogórski, *Studia*; Józef Mitkowski, "Uwagi o zaludnieniu Polski na początku panowania Kazimierza Wielkiego," *Roczniki Dziejów Społecznych i Gospodarczych* X (1948), 121-130; Witold Kula, "Stan i potrzeby badań nad demografią historyczną dawnej Polski (do początków XIX wieku)," *Roczniki Dziejów Społecznych i Gospodarczych* XII (1951), 23-106; Egone Vielrose, "Ludność Polski od X do XVIII wieku," *Kwartalnik Historii Kultury Materialnej* V (1957), 3-49.

[57] Gromnicki, *Świętopietrze*, 68; Ładogórski, *Studia*, 69-74; Tadeusz Ładogórski, "Spór o ocenę rachunków świętopietrza i liczebność zaludnienia Polski XIV wieku," *Kwartalnik Historii Kultury Materialnej* X (1962), 33–51.

[58] Gromnicki, *Świętopietrze*, 69.

[59] Benedictow, *The Black Death*, 383.

[60] Gromnicki, *Świętopietrze*, 54.

[61] *Monumenta Poloniae Vaticana*, vol. 2, ed. Jan Ptaśnik (Kraków, 1913), 81.

[62] Ibidem, 191-209; 238-255.

[63] Gromnicki, *Świętopietrze,* 67-80.

[64] Ładogórski, *Studia,* 113.

[65] Henryk Paszkiewicz, *Polityka ruska Kazimierza Wielkiego* (2ed., Kraków, 2002), 169-195; Paul W. Knoll, *The Rise of the Polish Monarchy. Piast Poland in East Central Europe, 1320-1370* (Chicago, 1972), 121-142, 143-152.

[66] *Najstarsze księgi i rachunki miasta Krakowa*, eds. Franciszek Piekosiński and Józef Szujski (Kraków, 1878); *Księgi ławnicze krakowskie 1365-1376 i 1390-1397*, eds. Stanisław Krzyżanowski (Kraków,1904); Agnieszka Bartoszewicz, *Urban Literacy in Late Medieval Poland* (Turnhout, 2017), 108-110.

[67] Bożena Wyrozumska, *Kancelaria miasta Krakowa w średniowieczu*, (Kraków,1995), 63.

[68] Ibidem.

[69] Jakub Wysmułek, *Testamenty mieszczan krakowskich (XIV-XV wiek)*, (Warszawa, 2015), 347-364.

[70] Adam Izdebski et al., "On the Use of Palynological Data in Economic History: New Methods and Application to Agricultural Output in Central Europe, 0-2000 AD," *Explorations in Economic History* LIX (2016), 17-29.

[71] Lee Mordechai, Merle Eisenberg, "Rejecting Catastrophe: The Case of the Justinianic Plague," *Past and Present* CCXXIV (2019), 3-50; Lee Mordechai et al., "The Justinianic Plague: An Inconsequential Pandemic?,"*Proceedings of the National Academy of Sciences of the United States of America,* CXVI (2019), 25546-25554; Merle Eisenberg, Lee Mordechai, "The Justinianic Plague and Global Pandemics: The Making of the Plague Concept", *American Historical Review* 125, 5 (2020: 1632-1667; Per Lagerås (ed.), *Environment, Society and the Black Death. An Interdisciplinary Approach to Late-medieval Crisis in Sweden* (Oxford, 2016).

[72] Izdebski et al., " On the Use of Palynological Data", 26.

[73] Katarzyna Marcisz et al., "Long-term Hydrological Dynamics and Fire History over the last 2000 years in CE Europe Reconstructed from a High-resolution Peat Archive," *Quaternary Science Reviews* CXI (2015), 138-152; S. Czerwiński *et al.*, "Znaczenie wspólnych badań historycznych i paleoekologicznych nad wpływem człowieka na środowisko. Przykład stanowiska Kazanie (Wschodnia Wielkopolska)," *Studia Geohistorica* VII (2019), 56-74; Mariusz Lamentowicz et al., "How Joannites's Economy Eradicated Primeval Forest and Created Anthroecosystems in Medieval Central Europe," *Scientific Reports* X (2020), 18775.

[74] Mitkowski, "Uwagi", 13; Kula, "Stan", 39; Vielrose, "Ludność", 48, Ładogórski, *Studia*, 134; Piotr Guzowski, "Stan i perspektywy badań nad liczbą ludności Polski w późnym średniowieczu i w początkach epoki nowożytnej", *Przeszłość Demograficzna Polski* XXVII (2015), 13.

[75] Irena Gieysztorowa, *Wstęp do demografii staropolskiej* (Warszawa, 1976), 146-184; Cezary Kuklo, *Demografia Rzeczypospolitej przedrozbiorowej* (Warszawa, 2009) 75-91.

[76] Kula*,* "Stan", 69; Vielrose, "Ludność"*,* 39; Guzowski, "Stan", 17.

[77] Irena Gieysztorowa, "Ludność", in Antoni Mączak (ed.), *Encyklopedia historii gospodarczej Polski do 1945 roku,* vol. 2 (Warszawa, 1981), 431.

[78] Andrew Hinde, *England's Population. A History since the Domesday Survey*, (London, 2003), 80.

[79] Christian Pfister, *Bevölkerungs-Geschichte und Historische Demographie* (München, 1994),11; Guliano Pinto, Eugenio Sonnino*,* "L'Italie", in Jean-Pierre Bardet and Jacques Dupâquier (eds.) *Histoire des populations de L'Europe* (Paris:Fayard 1997), 492; In the same volume data for other countries are given by: Etienne Helin, Ad van der Woude, "Les Pays-Bas", 411; Vincente Pérez Moreda, Robert L. Rowland, "La péninsule Ibérique", 463.

[80] Birabean, *Les hommes*, 17; Bruce M.S. Campbell, "The European Mortality Crises of 1346-52 and Advent of the Little Ice Age" ", in Dominik Collet and Maximilian Schuh (eds.), *Famines During the Little Ice Ages (1300-1800): socionatural entanglements in premodern societies* (Cham, 2017), 22.

[81] Ireneusz Ihnatowicz, Antoni Mączak, Benedykt Zientara, *Społeczeństwo polskie od X do XX wieku* (Warszawa 1979), 197.

[82] Benedictow, *The Black Death*, 221;Wilhelm Abel, Die Wüstungen des ausgehenden Mittelalters (Stuttgard, 1955), 100-101.

[83] Benedictow, *The Complete History*, 595; Wilhelm Abel, *Agrarkrisen und Agrarkonjunktur: Eine Geschichte de Land -und Ernährungswitschaft Mitteleuropas seit dem hohen Mittelalter* (Berlin, 1966), 58; Wilhelm Abel, *Agricultural Fluctuations in Europe from the Thirteenth to the Twentieth Centuries* (London 1980), 52.

[84] Julian Pelc, *Ceny w Krakowie w latach 1369-1600* (Lwów, 1935).

[85] Pelc, *Ceny*, 6-15, 80, 87.

[86] Janusz Kurtyka, *Odrodzone królestwo, Monarchia Władysława Łokietka i Kazimierza Wielkiego w świetle najnowszych badań* (Kraków, 2001), 115.

[87] Adam Rutkowski, "Objazdy i system rządzenia państwem przez Kazimierza Wielkiego," *Kwartalnik Historyczny* LXXXV (1978), 605-625; Antoni Gąsiorowski, "Itinerarium króla Kazimierza Wielkiego. Materiały 1333-1370," *Roczniki Historyczne* LXIV (1998),175-208.

[88] Stanisław Szczur, "Supliki Kazimierza Wielkiego," *Roczniki Historyczne* LIX (1993),67.

[89] Jarosław Widawski, "Mury obronne w Polsce Kazimierza Wielkiego," *Kwartalnik Historii Kultury Materialnej* XX (1972), 42.

[90] Jan Wroniszewski, "Król jako właściciel ziemski w średniowiecznej Polsce," in Andrzej Marzec and Maciej Wilamowski (eds.) *Król w Polsce XIV i XV wieku* (Kraków, 2003), 120.

[91] Sławomir Gawlas, "Polska Kazimierza Wielkiego a inne monarchie Europy Środkowej – możliwości i granice modernizacji władzy," in Marian Dygo et ali (eds.), *Modernizacja struktur władzy w warunkach opóźnienia. Europa Środkowa i Wschodnia na przełomie średniowiecza i czasów nowożytnych* (Warszawa, 1999), 27-28; Idem, "Król i stany w późnośredniowiecznej Europie Środkowo-Wschodniej wobec modernizacji państwa," in Marzec, Wilamowski (eds.), *Król w Polsce*,166-167; Andrzej Janeczek, "Ethnicity, Religious Disparity and the Formation of the Multicultural Society of Red Ruthenia in the Late Middle Ages," in Thomas Wünsch and Andrzej Janeczek, eds., *On the Frontier of Latin Europe. Integration and Segregation in Red Ruthenia, 1350-1600* (Warsaw, 2004), 15-45.

[92] Maria Bogucka, Henryk Samsonowicz, *Dzieje miast i mieszczaństwa w Polsce przedrozbiorowej* (Wrocław, 1986),86

[93] Karol Stefański, "Wsie na prawie niemieckim w Wielkopolsce w latach 1333-1370," *Roczniki Historyczne* XXXVII (1971), 1- 37.

[94] Jerzy Masłowski, "Kolonizacja wiejska na prawie niemieckim w województwach sieradzkim, łęczyckim, na Kujawach i w ziemi dobrzyńskiej," *Roczniki Historyczne* XIII (1937), 197-303; E. Dybek, "Lokacje na prawie niemieckim in cruda radice w południowej części województwa krakowskiego w latach 1334-1434," *Roczniki Humanistyczne* XLI (1993), 5-82.

[95] Konstanty Jan Hładyłowicz, *Zmiany krajobrazu i rozwój osadnictwa w Wielkopolsce od XIV do XIX wieku* (Lwów, 1932).

[96] Stanisław Gawęda, "Polityka fiskalna Kazimierza Wielkiego wobec kurii papieskiej," *Studia Historyczne* XIII (1970), 357-372; Jan Dudziak, *Dziesięcina papieska w Polsce średniowiecznej. Studium historyczno-prawne* (Lublin, 1974); Stanisław Szczur, *Annaty papieskie w Polsce w XIV wieku* (Kraków, 1998).

[97] Stanisław Szczur, "Papiestwo awiniońskie wobec odrodzonego Królestwa Polskiego w XIV wieku," in *Europa środkowa i wschodnia w polityce Piastów* (Toruń, 1997), 106-107.

[98] Yves Renouard, *Les Relationes des papes d'Avignon et des compagnies commerciales et bancaires de 1316 a 1378* (Paris,1941), 210.

[99] Marian Małowist, *Wschód a Zachód Europy w XIII-XVI wieku* (Warszawa, 2006), 39.

[100] Mengel, "A Plague on Bohemia? "; Andrea Kiss et al., "Food Crisis in Fourteenth-Century Hungary: Indicators, Causes and Case Studies", in Martin Bauch and Gerrit Jasper Schenk, *The Crisis of the 14$^{th}$ Century. Teleconnections between Environmental and Societal Change* (Berlin 2020), 100-129.

[101] Eduard Maur, "Příspěvek k demografické problematice předhusitských Čech 1346-1419," *AUC Philosophica et historica* I (1989), 7-72; Jaroslav Mezník, "Mory v Brně ve 14 století," *Miediaevalia Historica Bohemica* III (1993) 225-234; Eduard Maur, "Obzvatelstvo českých zemí ve středověku," in Ludmila Fialova et al. (ed.) *Dějiny obzvatelstva českých zemí* (Praha 1996), 35-74; Możejko, "Zarazy w średniowiecznym Gdańsku".

[102] Monica H. Green, "Editor's Introduction," in *Pandemic Disease in the Medieval World: Rethinking the Black Death* (Croydon, 2014), 9-26; John Haldon et al., "History Meets Palaeoscience: Concilience and Collaboration in Studying Past Societal Responses to Environmental Change," *Proceedings of the National Academy of Sciences of the United States of America,* CXV (2018), 3210-3218.